\documentclass[preprint,12pt]{elsarticle}

\usepackage{amssymb}
\usepackage{amsmath}
\usepackage{booktabs} 
\usepackage{multirow}
\usepackage{tikz}
\usepackage{pgfplots}
\usepackage{float}
\pgfplotsset{compat=1.18}
\usepackage{xcolor} 
\usepackage{subcaption}
\usepackage{graphicx}
\usepackage[ruled,vlined,linesnumbered]{algorithm2e}

\pgfplotsset{compat=newest}
\definecolor{IGABlue}{RGB}{60,120,210} 
\definecolor{OrigColor}{RGB}{255,190,140}  
\definecolor{RDAColor}{RGB}{150,220,170}   
\definecolor{SACNColor}{RGB}{190,190,255}

\journal{Physica A}

\begin{document}

\begin{frontmatter}



\title{IGA-LWP: An Iterative Gradient-based Adversarial Attack for Link Weight Prediction} 


            \author[label1]{Cunlai Pu\corref{cor1}}   
            \cortext[cor1]{Corresponding author}
            \ead{pucunlai@njust.edu.cn}
            \author[label1]{Xingyu Gao}
            \author[label1]{Jinbi Liang}
            \author[label1]{Jianhui Guo}
            \author[label1]{Xiangbo Shu}
            \author[label2]{Yongxiang Xia}
            \author[label3]{and Rajput Ramiz Sharafat}
            
            \affiliation[label1]{organization={School of Computer Science and Engineering, Nanjing University of Science and Technology},
            	city={Nanjing},
            	postcode={210094}, 
            	state={Jiangsu},
            	country={China}}
            
            \affiliation[label2]{organization={School of Communication Engineering, Hangzhou Dianzi University},
            	city={Hangzhou},
            	postcode={310018}, 
            	state={Zhejiang},
            	country={China}}
            
            \affiliation[label3]{organization={School of Computer Science and
            		Technology, University of Science and Technology of China},
         	city={Hefei},
         	postcode={230026}, 
         	state={Anhui},
         	country={China}}

\begin{abstract}
Link weight prediction extends classical link prediction by estimating the strength of interactions rather than merely their existence, and it underpins a wide range of applications such as traffic engineering, social recommendation, and scientific collaboration analysis. However, the robustness of link weight prediction against adversarial perturbations remains largely unexplored.
In this paper, we formalize the link weight prediction attack problem as an optimization task that aims to maximize the prediction error on a set of target links by adversarially manipulating the weight values of a limited number of links. Based on this formulation, we propose an iterative gradient-based attack framework for link weight prediction, termed IGA-LWP. By employing a self-attention–enhanced graph autoencoder as a surrogate predictor, IGA-LWP leverages backpropagated gradients to iteratively identify and perturb a small subset of links.
Extensive experiments on four real-world weighted networks demonstrate that IGA-LWP significantly degrades prediction accuracy on target links compared with baseline methods. Moreover, the adversarial networks generated by IGA-LWP exhibit strong transferability across several representative link weight prediction models. These findings expose a fundamental vulnerability in weighted network inference and highlight the need for developing robust link weight prediction methods.
\end{abstract}




\end{frontmatter}



\section{Introduction}
\label{sec_Intro}

Many real-world systems can be naturally modeled as graphs, where vertices represent entities and links encode interactions between them. Examples include social networks \cite{singh2024social}, biological networks \cite{raval2013introduction}, communication networks \cite{monge2003theories}, and smart grids \cite{powell2024smart}. In many of these systems, links are not merely present or absent: they carry weights that quantify the strength, frequency, or capacity of interactions, such as message volume between users, binding affinity between proteins, or power flow along transmission lines \cite{newman2018networks,lewis2011network,posfai2016network}. Link prediction \cite{lu2011link,zhou2021progresses,liben2003link} in its classical form answers a binary question—whether a link between two nodes will exist—whereas link weight prediction \cite{lu2010link,fu2018link,kumar2016edge} aims to estimate the value of the link weight. This finer-grained task enables more precise network analysis and directly supports downstream decision making. For example, in social networks, link weight prediction can capture the frequency or intimacy of user interactions and thus improve friend recommendation; in recommender systems, it can model user preference intensity for items and enhance recommendation accuracy and satisfaction; in biological networks, it helps analyze the strength of protein–protein interactions to support drug discovery and disease research; and in transportation and communication networks, predicting traffic flow or congestion levels on links can guide traffic management, routing, and capacity planning. Accordingly, link weight prediction has become a key tool for understanding and optimizing complex networked systems.

The rapid development and broad deployment of deep learning have significantly advanced performance in computer vision, natural language processing, and many other domains \cite{goodfellow2016deep,dong2021survey}. At the same time, deep models have been shown to suffer from serious security and robustness issues \cite{akhtar2018threat}. A prominent example is the adversarial attack phenomenon, where carefully crafted, small perturbations added to the input can cause a model to produce highly erroneous predictions \cite{yuan2019adversarial}. Extending deep learning to graph-structured data, Graph Neural Networks (GNNs) and related architectures have substantially improved the performance of a variety of graph analysis tasks by learning non-linear, hierarchical representations that capture latent node and link features \cite{wu2020comprehensive,corso2024graph}. However, GNN-based methods inherit many of the vulnerabilities of deep models in Euclidean domains: their predictions can be highly sensitive to subtle, structured changes in the input graph.

Motivated by these concerns, a growing body of work has studied adversarial attacks in graph analysis \cite{xu2021robustness,sun2022adversarial}. Nagaraja \cite{nagaraja2010impact} first investigated community deception attacks against community detection algorithms, highlighting privacy risks in graph analytics. Zügner  et al. proposed NETTACK \cite{zugner2018adversarial}, an iterative attack that perturbs graph structure and node attributes based on the change in prediction confidence, and demonstrated its effectiveness in degrading node classification performance. For link prediction, Chen et al. \cite{chen2020link} developed a graph autoencoder-based attack algorithm to generate adversarial graphs that degrade prediction performance.
Zheleva and Getoor introduced link re-identification attacks \cite{zheleva2007preserving}, arguing that link prediction itself can be viewed as an attack because it may expose sensitive relationships in released graph data. Other works have targeted specific graph mining algorithms, such as fast gradient attacks (FGA)  on node embedding \cite{chen2018fast}. Collectively, these studies show that graph-based learning methods, despite their strong predictive performance, can be surprisingly fragile under carefully designed perturbations.

In contrast, adversarial attacks on link weight prediction have received much less attention, even though their importance should not be underestimated. Studying attacks on link weight prediction offers a principled way to evaluate the robustness of these algorithms, expose potential vulnerabilities, and guide the design of more secure models. On the other hand, controlled perturbations of link weights provide a complementary perspective for privacy protection: by deliberately adjusting weights, one can prevent sensitive information from being predicted by adversaries. In network security and privacy, such attacks can both highlight risks and inspire defense strategies.

These observations motivate us to investigate  adversarial attacks on link weight prediction in weighted graphs. We focus on the setting where an attacker aims to hide or distort the weights of specific target links by modifying only a small number of  weight values of other links in the underlying graph. The attacker may have complete or incomplete knowledge of the graph as prior information.
The main contributions of our work are summarized as follows:
\begin{itemize}
	\item We formally define the adversarial attack problem for link weight prediction in weighted graphs as a constrained optimization task that maximizes the prediction error on a set of target links under a strict budget on the number of perturbed link weights. The formulation provides a general framework for analyzing the 	robustness of link weight predictors.
	\item  We propose  IGA-LWP, an iterative gradient-based attack model  on link weight prediction. This model uses a self-attention enhanced graph auto-encoder (SEA) \cite{liu2023self} as a surrogate model, and leverages backpropagated gradients to identify a constraint number of  influential links to attack. This model can be adapted to both global attacks (manipulating arbitrary links in the whole graph) and local attacks (restricting perturbations to links incident to the endpoints of the target link).	\item   Experiments on real-world weighted networks show that IGA-LWP significantly degrades prediction accuracy on target links compared with baseline methods, and that the adversarial graphs produced by IGA-LWP substantially degrade the performance of diverse link weight predictors, demonstrating strong transferability and revealing a fundamental robustness issue for link weight prediction in weighted graphs.
\end{itemize}

The remainder of this paper is organized as follows. Section 2 introduces the problem of link weight prediction attacks. Section 3 details our proposed method, IGA-LWP. Section 4 presents the performance evaluation of the proposed method. Section 5 concludes the paper.

\section{Problem formulation}
\label{sec_PF}
We consider an undirected weighted network represented by a triple
\(
\mathcal{G} = (\mathcal{V}, \mathcal{E}, \mathcal{W}),
\)
where $\mathcal{V}$ is the set of nodes, $\mathcal{E}$ is the set of links, and $\mathcal{W}$ is the set of link weights.  Let $\mathcal{E}^* \subseteq \mathcal{E}$ be a set of links whose weights $\mathcal{W}^* \subseteq \mathcal{W}$ are missing or unavailable. Given the observed network
\(
\mathcal{G}_o = (\mathcal{V}, \mathcal{E}, \mathcal{W} \setminus \mathcal{W}^*),
\)
the goal of link weight prediction is to recover the missing weights $\mathcal{W}^*$ as accurately as possible.

The observed network is represented by an adjacency matrix
\(
A \in \{0,1\}^{|\mathcal{V}|\times |\mathcal{V}|}
\), 
where entry $a_{uv}=1$ if there is a link between nodes $u$ and $v$, otherwise  $a_{uv}=0$. It can  also be represented by  a weighted adjacency matrix
\(
W \in \mathbb{R}_{+}^{|\mathcal{V}|\times |\mathcal{V}|},
\)
where entry $w_{u,v}$ is the weight of the link between nodes $u$ and $v$; $w_{u,v}=0$ indicates that there is no link or the weight is missing. We consider undirected graphs, hence $w_{u,v}=w_{v,u}$ and self-loops are not included, i.e.,  $a_{u,u}=0$ and $w_{u,u}=0$.
To avoid the influence of large weight ranges on the prediction performance, link weights are  normalized as
\begin{equation}
w_{\text{new}} = e^{-\frac{1}{w_{\text{old}}}},
\end{equation}
where the resulting  weight values fall in the   interval $(0,1)$.

\label{subsec2-pf}
 We denote by $\Delta \mathcal{W}_{\beta}$ a small perturbation on the weight set $\mathcal{W}$, and thus obtain a new weight set $\hat{\mathcal{W}} = \mathcal{W} + \Delta \mathcal{W}_{\beta}$  corresponding to the  generated adversarial graph $\hat{\mathcal{G}}$, where nodes and links are the same as the original graph, yet the link weights are different. 
Each element of $\Delta \mathcal{W}_{\beta}$ can be a positive value, indicating an increase in the corresponding link weight, or a negative value, meaning a decrease.

Let $f$ be a link weight prediction method, and $\mathcal{E}_t \subseteq \mathcal{E}$ be a set of target links with corresponding weight set  $\mathcal{W}_t$. The aim of adversarial attack is to make the predicted weight set $\hat{\mathcal{W}}_t = f(\hat{\mathcal{G}},\mathcal{E}_t)$ significantly deviates from $\mathcal{W}_t$. 

Formally, for a given graph $\mathcal{G}$ and a target link set $\mathcal{E}_t$, a link weight prediction adversarial attack seeks an adversarial graph $\hat{\mathcal{G}}$ that maximizes the discrepancy between predicted and true weight values:
\begin{equation}
\max_{\hat{\mathcal{G}}} \; D\big(f(\hat{\mathcal{G}}, \mathcal{E}_t),\, \mathcal{W}_t \big)
\quad \text{s.t.} \quad |\Delta \mathcal{E}| \le m,
\end{equation}
where $D(\cdot,\cdot)$ is a discrepancy measure, $|\Delta \mathcal{E}|$ is the number of links whose weights are modified, and $m$ is an upper bound on the number of perturbed links.

\section{Method}
\label{sec_Methods}
\label{subsec_frame}
\begin{figure}[h]
\centering
\includegraphics[width=\textwidth]{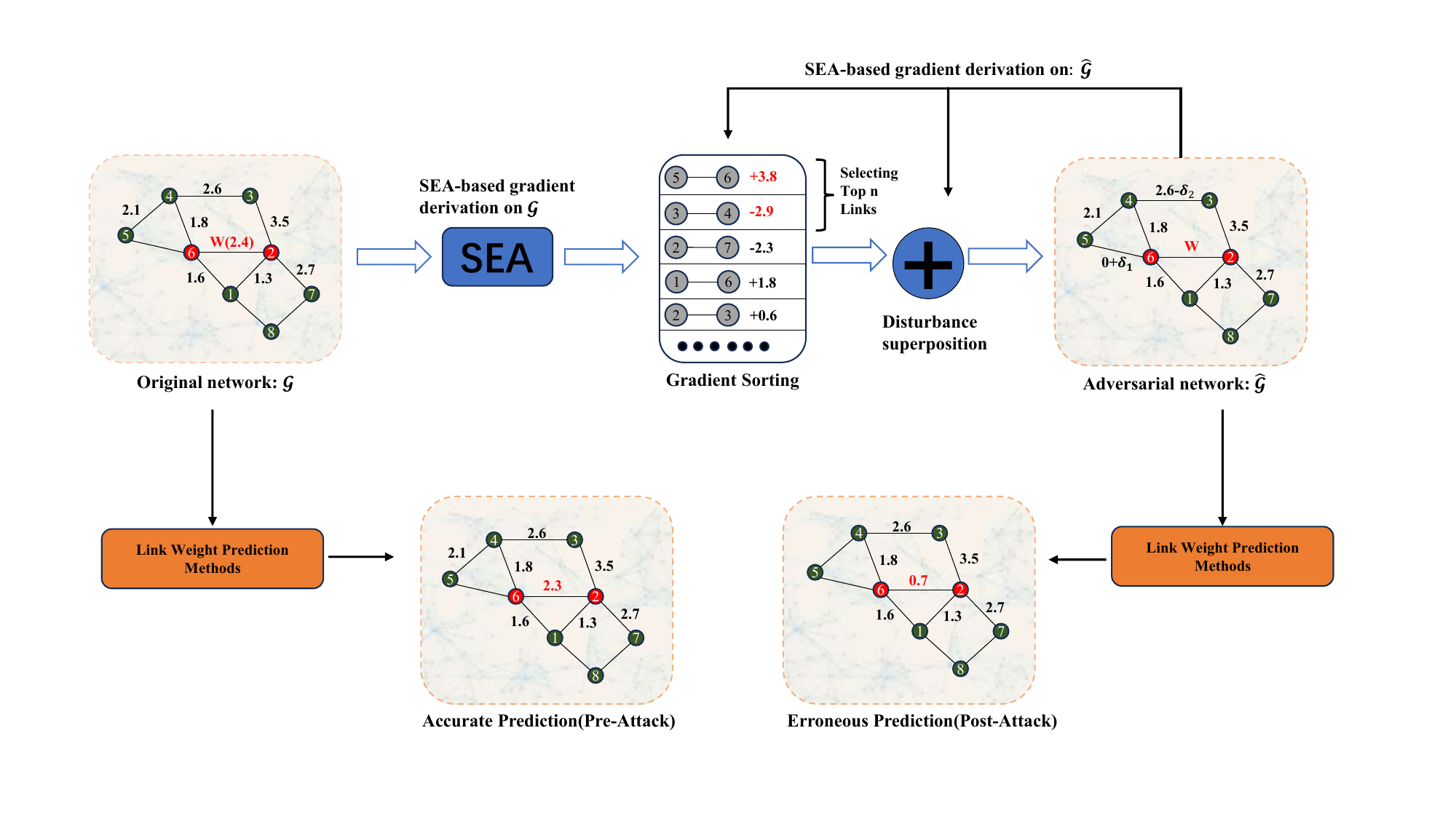}
\caption{Framework of the IGA-LWP model for adversarial link weight prediction.
	This model includes SEA-based gradient derivation and gradient sorting to select key links for perturbations. Perturbations are then superimposed to construct an adversarial network, which leads to erroneous link weight predictions.}
\label{frame}
\end{figure}

The overall framework of IGA-LWP is illustrated in Fig. \ref{frame}. We first select a link as the attack target, with the goal of substantially reducing the probability that its weight can be accurately predicted by  link weight prediction models.  We design an attack loss function for this target link, and  compute the gradient matrix of the loss function with respect to the weight matrix, and use this gradient information to generate the corresponding adversarial graph iteratively.

\subsection{Link weight prediction model}
\label{subsec1}
We adopt the self-attention enhanced graph auto-encoder (SEA) \cite{liu2023self} as the surrogate model for gradient-based attacks. SEA is a link-level auto-encoder composed of a link encoder and a regression decoder.
To capture nonlinear deep graph features while considering both first-order neighborhood and second-order structural information, each node $u$ is initially represented  as
\begin{equation}
    h_u = [W x_u \;\| \; A^2 x_u],
\end{equation}
where  $x_u$  is a one-hot vector,  i.e., a column of the $|\mathcal{V}|\times|\mathcal{V}|$ identity matrix,  $\|$ denotes concatenation and $A^2$ is the second-order adjacency matrix whose entries count common neighbors between node pairs.

 SEA employs a Graph Attention Network (GAT) \cite{velickovic2017graph} to aggregate information. For a neighbor $k$ of node $u$, the attention coefficient $\alpha_{u,k}$ measuring the importance of node $k$ to node $u$ is computed as
\begin{equation}
\alpha_{u,k} = \frac{\exp\big(\text{LeakyReLU}(\gamma^\top \rho_{u,k})\big)}
{\sum_{j \in \mathcal{N}_u} \exp\big(\text{LeakyReLU}(\gamma^\top \rho_{u,j})\big)},
\end{equation}
where $\rho_{u,k}$ is an affine transformation of $[h_u \| h_k]$, $\gamma$ is a learnable parameter vector, $\mathcal{N}_u$ is the neighbor set of node $u$, and the LeakyReLU has negative slope $0.2$.

 Based on learned attention coefficients related to nodes $u$ and $v$, the aggregated embedding for link $(u,v)$ is
\begin{equation}
B_{u,v} = \text{LeakyReLU} \Big( \Omega_3 \big[ \sum_{k\in \mathcal{N}_u}\alpha_{u,k} h_k \;\| \; \sum_{j\in \mathcal{N}_v}\alpha_{v,j} h_j \big] \Big),
\end{equation}
where $\Omega_3$ is a learnable matrix.
The decoder maps the embedding to the predicted weight of  link  $(u,v)$:
\begin{equation}
    w'_{u,v} = \sigma(\Theta^\top B_{u,v}),
\end{equation}
where $\Theta$ is a parameter vector and $\sigma(\cdot)$ is the sigmoid function.

To learn optimal link embeddings and minimize  prediction error, SEA minimizes the loss
\begin{equation}
\mathcal{L} = \sum_{u\ne v} a_{u,v} (w_{u,v}-w'_{u,v})^2
+ \nu \sum_{u\ne v} a_{u,v} \|B_{u,v}-B_{v,u}\|_2^2
+ \mathcal{L}_{\text{reg}},
\end{equation}
where the second term enforces a symmetry  regularization, encouraging the embeddings of the two directions of a node pair to be close, and $\mathcal{L}_{\text{reg}}$ is an $\ell_2$ regularization on the parameters to prevent overfitting.
\subsection{Gradient extraction for target links}
\label{subsec2-me}
In the training of SEA, the loss is computed over all observed links. 
For IGA-LWP, however, we focus on a single target link $(u,v)$. We define a target loss
\begin{equation}
    \mathcal{L}_t = (w_{u,v}-w'_{u,v})^2,
\end{equation}
where $w_{u,v}$ is the true weight of the target link and $w'_{u,v}$ is the prediction of SEA.
The gradient of the target loss with respect to the weight matrix $W$ can be obtained via the chain rule:
\begin{equation}
    g_{ij} = \frac{\partial \mathcal{L}_t}{\partial W_{ij}}.
\end{equation}

 Since SEA   does not enforce symmetry  of the  gradient matrix,  we symmetrize it and keep its upper triangular part:
\begin{equation}
\hat{g}_{ij} =
\begin{cases}
\frac{1}{2}(g_{ij}+g_{ji}), & i<j,\\
0, & \text{otherwise}.
\end{cases}
\end{equation}
\subsection{Iterative generation of adversarial graphs}

In standard SEA training, we minimize the global reconstruction loss to obtain good predictions. In the adversarial setting, we instead maximize the target loss $\mathcal{L}_t$, thereby inducing large prediction errors on the target link.

 For links not present in the graph ($a_{ij}=0$), there is no weight to adjust. Thus gradient analysis and perturbation are performed only on existing links with $a_{ij}=1$.
For a link weight $w_{ij}$, the sign of its gradient indicates how it should be modified:
\begin{itemize}
    \item If $\dfrac{\partial \mathcal{L}_t}{\partial w_{ij}}>0$, increasing $w_{ij}$ will increase the target loss; we update
    \begin{equation}
        w_{ij} \leftarrow w_{ij} + \eta \bigg|\frac{\partial \mathcal{L}_t}{\partial w_{ij}}\bigg|.
    \end{equation}
    \item If $\dfrac{\partial \mathcal{L}_t}{\partial w_{ij}}<0$, decreasing $w_{ij}$ will increase the target loss; we update
    \begin{equation}
        w_{ij} \leftarrow w_{ij} - \eta \bigg|\frac{\partial \mathcal{L}_t}{\partial w_{ij}}\bigg|.
    \end{equation}
\end{itemize}
Here $\eta$ is a learning rate controlling the perturbation magnitude, and $\left|\dfrac{\partial \mathcal{L}_t}{\partial w_{ij}}\right|$ measures how strongly the link weight affects the target loss.

At each iteration, we select $n$ edges with the largest gradient magnitudes $\left|\dfrac{\partial \mathcal{L}_t}{\partial w_{ij}}\right|$, and update their weights as above. Repeating this process for $K$ iterations yields the final adversarial graph.
 
The pseudocode of IGA-LWP is a combination of Algorithms 1 and 2.
\label{app1}
\begin{algorithm}[htbp]
	\caption{Adversarial Graph Generator}
	\label{alg:adv-graph-generator-en}
	\KwIn{Original graph $\mathcal{G}$, number of iterations $K$, number of weights to modify per iteration $n$}
	\KwOut{Adversarial graph $\hat{\mathcal{G}}$}
	
	Train a link-weight prediction model (e.g., SEA) on graph $\mathcal{G}$\;
	Initialize the weight matrix of the adversarial graph as
	$\hat{W}^{0}=W$ (where $W$ is the weight matrix of the original graph)\;
	\For{$h=1$ \KwTo $K$}{
		Compute the gradient matrix $g^{h-1}$ based on the current weight matrix $\hat{W}^{h-1}$\;
		Symmetrize the gradient matrix $g^{h-1}$ to obtain $\hat{g}^{h-1}$\;
		$P \leftarrow \textsc{WeightPerturbationGenerator}
		\big(\hat{W}^{h-1}, \hat{g}^{h-1}, n\big)$\;
		$\hat{W}^{h} \leftarrow \hat{W}^{h-1} + P$\;
	}
	Return the adversarial graph $\hat{\mathcal{G}}$ whose weight matrix is $\hat{W}^{K}$\;
\end{algorithm}

\begin{algorithm}[H]
	\caption{ Weight Perturbation Generator}
	\label{alg:construct-weight-perturbation-en}
	\KwIn{Adjacency matrix $A$, weight matrix $W$, symmetrized gradient matrix $\hat{g}^{h-1}$, number of weights to modify $n$}
	\KwOut{Weight perturbation matrix $P$}
	
	Initialize the weight perturbation matrix $P$ as a zero matrix with the same size as $W$\;
	\For{$h=1$ \KwTo $n$}{
		Find the position $(i,j)$ of the element with the largest absolute value in $\hat{g}^{h-1}$\;
		\If{$\hat{g}^{h-1}_{ij} > 0$ \textbf{and} $A_{ij}=1$}{
			$P_{ij} \leftarrow +\epsilon$, where $\epsilon > 0$ is the predefined perturbation magnitude\;
		}
		\ElseIf{$\hat{g}^{h-1}_{ij} < 0$ \textbf{and} $A_{ij}=1$}{
			$P_{ij} \leftarrow -\epsilon$\;
		}
		\Else{
			\textbf{continue}\;
		}
	}
	$P \leftarrow P + P^{\mathsf T}$, where $P^{\mathsf T}$ is the transpose of $P$\;
	Return the weight perturbation matrix $P$\;
\end{algorithm}

\subsection{Global and local attacks}

In IGA-LWP, adversarial graphs are generated using SEA as a surrogate, which corresponds to a typical white-box attack on SEA: all model parameters and gradients are available. However, the attack capability can be constrained by the attacker's access to the graph.

In the global attack scenario, the attacker can freely choose any link in the network for weight perturbation, limited only by the total number of perturbed links. This corresponds to a high-privilege attacker, such as a data publisher, who wishes to hide sensitive information or relationships by slightly adjusting link weights while preserving the overall utility of the network \cite{fung2010privacy}.

In the local attack scenario, the attacker can only modify the weights of links connected to one endpoint of the target link and cannot change the weights of distant links. This reflects more realistic situations where the attacker has limited access to the graph. For example, a user in a recommender system who can only manipulate interactions related to their own accounts, but not the entire network \cite{fang2018poisoning}.

These two scenarios model different levels of attacker knowledge and privileges and are used in our work.

\subsection{Transferability of   adversarial attacks}

Transferable adversarial attacks aim to successfully compromise prediction models without accessing their internal details \cite{alvarez2023exploring}. Specifically, an attacker can generate adversarial graphs using one model and apply them to other unknown link weight prediction methods to observe changes in prediction results and evaluate the effectiveness of the attack.

Adversarial graphs generated based on the IGA-LWP method capture critical structural information within the graph, granting the adversarial perturbations a certain degree of generality. As a result, these adversarial graphs remain effective against other prediction models such as DeepWalk \cite{perozzi2014deepwalk}, Node2Vec \cite{grover2016node2vec} and GCN \cite{chen2020simple}.

\section{Performance evaluation}

\subsection{Datasets}

We evaluate the proposed model on four weighted networks of different types and scales. 
Their basic statistics are summarized in Table~\ref{tab:dataset_feature}. 
A brief description of each network is given below.
\begin{itemize}
	\item \textbf{Neural-net}~\cite{neural}: a neural network of C. elegans, where nodes represent neurons and links correspond to synaptic or gap junction connections. Link weights indicate the number of interactions between neurons.
		\item \textbf{C. elegans}~\cite{cgans}: a metabolic network of C. elegans, where links represent interactions between metabolites, and weights reflect the multiplicity of interactions.
		\item \textbf{Netscience}~\cite{newman}: the largest connected component of a coauthorship network in network science. Link weights are computed based on coauthored papers and coauthor information.
		\item \textbf{UC-net}~\cite{ucnet}: a communication network of an online student community at the University of California, Irvine, where users are nodes and directed links represent message flows. We remove link directions and aggregate multiple links between two nodes; the link weight represents the number of messages exchanged between nodes.
\end{itemize}

\begin{table}[htbp]
  \centering
  \begin{tabular}{lrrrr}
    \toprule
    \textbf{Dataset} & \textbf{\#Nodes} & \textbf{\#Edges} & \textbf{Weight range} & \textbf{Type} \\
    \midrule
    Neural-net  &   296   &  2\,137 & $[1,72]$        & Biology      \\
    C. elegans   &   453   &  2\,025 & $[1,114]$       & Biology      \\
    Netscience  &   575   &  1\,028 & $[0.0526,2.5]$  & Coauthorship \\
    UC-net      & 1\,899  & 13\,828 & $[1,184]$       & Social       \\
    \bottomrule
  \end{tabular}
  \caption{Basic topological features of the weighted networks.}
  \label{tab:dataset_feature}
\end{table}
\subsection{Evaluation metrics}

We use two standard metrics for link weight prediction, which are as follows.

\paragraph{Pearson Correlation Coefficient (PCC)} PCC measures linear correlation between predicted and true weights:
\begin{equation}
	\text{PCC} = 
	\frac{\sum_{i=1}^n (\hat{y}_i-\bar{\hat{y}})(y_i-\bar{y})}
	{\sqrt{\sum_{i=1}^n (\hat{y}_i-\bar{\hat{y}})^2}
		\sqrt{\sum_{i=1}^n (y_i-\bar{y})^2}},
\end{equation}
where $n$ is the number of samples, $y_i$ and $\hat{y}_i$ denote the true and predicted link weights, and $\bar{y}$ and $\bar{\hat{y}}$ are their corresponding means.

\paragraph{Root Mean Squared Error (RMSE)} RMSE measures the average magnitude of prediction errors:
\begin{equation}
\text{RMSE} = \sqrt{\frac{1}{n} \sum_{i=1}^n (y_i - \hat{y}_i)^2},
\end{equation}
where $y_i$ and $\hat{y}_i$ are true and predicted weights.

The goal of adversarial attack is to decrease PCC and increase RMSE, i.e., make predictions less correlated with and further away from the true weights.

\subsection{Baseline attack methods}

We compare IGA-LWP with two baselines:
\begin{itemize}
    \item RDA (Random Attack): randomly selects a given number of links and perturbs their weights. It does not use any structural or gradient information and serves as a simple baseline.
\item SA-CN (Similarity-based Attack–Common Neighbors): selects links whose endpoints have a large number of common neighbors \cite{newman2001clustering} and introduces small perturbations to their weights. Perturbing such important links can effectively disrupt structurally coherent regions of the network and degrade link weight prediction performance.
\end{itemize}
Both baselines use the same perturbation budget and magnitude as IGA-LWP for fair comparison.
\subsection{Experimental setup}

The experiments are conducted in the following software and hardware environment: Windows 11, Python 3.10, PyTorch 1.12.1, Intel i9-12900H CPU (2.50 GHz), Nvidia RTX 3070 GPU, and 16 GB RAM.

Following the experimental setup in SEA, we randomly select 10\% of links as the test set and use the remaining 90\% as the observed network for training SEA. The trained SEA model serves as the target model for the adversarial attack. From the test set, we randomly select 10 target links  for attack. To reduce variance due to randomness, all reported results are averaged over 10 independent runs.

We define the perturbation budget based on the degree of the target link. Let \(k_t\) denote the sum of degrees of the two endpoints of the target link. For baseline attacks, we set the number of perturbed links to \(0.5 k_t\) and use a perturbation magnitude \(\delta_{ij} = \alpha w_{ij}\), where \(\alpha\) is a scalar factor. For IGA-LWP, we set \(n=1\) (only one link updated per iteration) and \(K = 0.5 k_t\). Thus, the total number of modified links is \(n \times K\), ensuring that IGA-LWP uses the same perturbation budget as the baseline methods.

\subsection{Comparison  of different methods under global and local attacks}

We evaluate the attack performance of IGA-LWP, RDA, and SA-CN against the prediction model SEA under both global and local attack settings. Tables \ref{tb_global} and \ref{tb_local} respectively report RMSE and PCC before and after attacks on the four datasets. The original SEA model achieves low RMSE and high PCC on all datasets, demonstrating its strong prediction performance.

Under global attack, IGA-LWP dramatically increases RMSE and reduces PCC for all datasets. In some cases, PCC even flips from positive to negative, indicating that predictions become anticorrelated with the true weights. In contrast, RDA and SA-CN lead to only minor changes in both metrics.

\begin{table}[]
\centering
\scriptsize
\begin{tabular}{@{}ccccccccc@{}}
\toprule
\multirow{2}{*}{Datasets} & \multicolumn{4}{c}{RMSE}                  & \multicolumn{4}{c}{PCC}                     \\
                          & ORGIN  & RDA    & SA-CN  & IGA-LWP         & ORGIN  & RDA    & SA-CN  & IGA-LWP          \\ \midrule
Neural                    & 0.1896 & 0.1904 & 0.1916 & \textbf{0.3207} & 0.4249 & 0.4103 & 0.4127 & \textbf{-0.1180} \\
C. elegans                 & 0.1002 & 0.1002 & 0.1003 & \textbf{0.2100} & 0.6552 & 0.6551 & 0.6507 & \textbf{-0.1351} \\
NetScience                & 0.0697 & 0.0698 & 0.0699 & \textbf{0.1055} & 0.7940 & 0.7939 & 0.7939 & \textbf{0.6070}  \\
UCsocial                  & 0.1874 & 0.1896 & 0.1896 & \textbf{0.3127} & 0.5204 & 0.4940 & 0.4943 & \textbf{-0.2223} \\ \bottomrule
\end{tabular}
\caption{Results of global attacks on  SEA  under different attack methods.}
\label{tb_global}
\end{table}

\begin{table}[]
\centering
\scriptsize
\begin{tabular}{@{}ccccccccc@{}}
\toprule
\multirow{2}{*}{Datasets} & \multicolumn{4}{c}{RMSE}                   & \multicolumn{4}{c}{PCC}                     \\
                          & ORGIN  & RDA    & SA-CN  & IGA-LWP         & ORGIN  & RDA    & SA-CN  & IGA-LWP          \\ \midrule
Neural                    & 0.1896 & 0.1954 & 0.1982 & \textbf{0.3013} & 0.4249 & 0.3621 & 0.3342 & \textbf{-0.1089} \\
C. elegans                  & 0.1002 & 0.1083 & 0.1036 & \textbf{0.1837} & 0.6552 & 0.5688 & 0.6347 & \textbf{0.0388}  \\
NetScience                & 0.0697 & 0.0805 & 0.0817 & \textbf{0.1092} & 0.7940 & 0.6311 & 0.6166 & \textbf{0.5567}  \\
UCsocial                  & 0.1874 & 0.2003 & 0.2006 & \textbf{0.3093} & 0.5204 & 0.4617 & 0.3567 & \textbf{-0.1156} \\ \bottomrule
\end{tabular}
\caption{Results of local attacks on  SEA  under different attack methods.}
\label{tb_local}
\end{table}

Under local attack, IGA-LWP still achieves the best attack performance by significantly degrading SEA's predictions while perturbing only links adjacent to the target link. RDA and SA-CN exhibit limited effectiveness; in some datasets, their attacks do not substantially affect PCC or RMSE. These results confirm that, even under local constraints, gradients extracted from SEA provide accurate directions for generating highly effective perturbations.

For IGA-LWP, global attacks generally outperform local attacks on datasets in which links with large gradient magnitudes are distributed throughout the graph (e.g., Neural-net, C. elegans, UC-net). However, on Netscience, high-gradient links tend to concentrate near the target link, making local attacks more targeted and competitive. For RDA, local attacks sometimes outperform global ones; this is because random global perturbations are more likely to affect unimportant links compared to local random perturbations. SA-CN’s performance depends heavily on the clustering structure and may be limited in sparse networks or those exhibiting random-like topology.

\subsection{Effect of different perturbation ratios on attack performance}
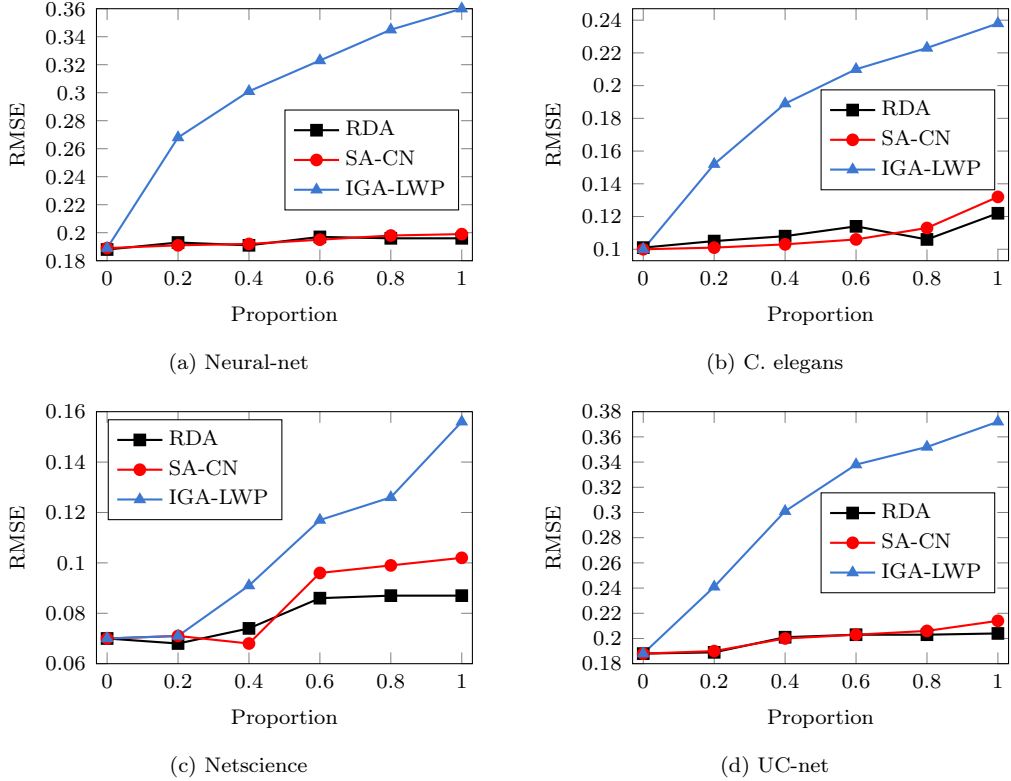
\begin{figure}[H]
    \centering

    \begin{subfigure}[t]{0.48\textwidth}
        \centering
        \begin{tikzpicture}
        \begin{axis}[
            width=\textwidth,
            height=0.75\textwidth,
            xlabel={Proportion},
            ylabel={RMSE},
            xmin=0, xmax=1.0,
            ymin=0.18, ymax=0.36,
            enlarge x limits=0.03,
            xtick={0,0.2,0.4,0.6,0.8,1.0},
            ytick={0.18,0.20,0.22,0.24,0.26,0.28,0.30,0.32,0.34,0.36},
            scaled y ticks=false,
            yticklabel style={
                /pgf/number format/fixed,
                /pgf/number format/precision=2
            },
            axis lines=box,
            tick align=inside,
            legend style={
                draw=black,
                fill=white,
                at={(0.5,0.6)},
                anchor=north west
            },
            font=\scriptsize,
            legend cell align=left
        ]
            \addplot[
                color=black,
                mark=square*,
                thick
            ] coordinates {
                (0.0,0.188)
                (0.2,0.193)
                (0.4,0.191)
                (0.6,0.197)
                (0.8,0.196)
                (1.0,0.196)
            };
            \addlegendentry{RDA}

            \addplot[
                color=red,
                mark=*,
                thick
            ] coordinates {
                (0.0,0.189)
                (0.2,0.191)
                (0.4,0.192)
                (0.6,0.195)
                (0.8,0.198)
                (1.0,0.199)
            };
            \addlegendentry{SA-CN}

            \addplot[
                color=IGABlue,
                mark=triangle*,
                thick
            ] coordinates {
                (0.0,0.189)
                (0.2,0.268)
                (0.4,0.301)
                (0.6,0.323)
                (0.8,0.345)
                (1.0,0.360)
            };
            \addlegendentry{IGA-LWP}
        \end{axis}
        \end{tikzpicture}
        \caption{Neural-net}
        \label{fig:sub-neural}
    \end{subfigure}
    \hfill
    \begin{subfigure}[t]{0.48\textwidth}
        \centering
            \begin{tikzpicture}
            \begin{axis}[
                width=\textwidth,
                height=0.75\textwidth,
                xlabel={Proportion},
                ylabel={RMSE},
                xmin=0, xmax=1.0,
                ymin=0.10, ymax=0.24,
                enlarge x limits=0.03,   
                enlarge y limits=0.05,   
                xtick={0,0.2,0.4,0.6,0.8,1.0},
                ytick={0.10,0.12,0.14,0.16,0.18,0.20,0.22,0.24},
                scaled y ticks=false,
                yticklabel style={
                    /pgf/number format/fixed,
                    /pgf/number format/precision=2
                },
                axis lines=box,
                tick align=inside,      
                legend style={
                    draw=black,
                    fill=white,
                    at={(0.5,0.68)},
                    anchor=north west
                },
                font=\scriptsize,
                legend cell align=left
            ]
        
                \addplot[
                    color=black,
                    mark=square*,
                    thick
                ] coordinates {
                    (0.0,0.101)
                    (0.2,0.105)
                    (0.4,0.108)
                    (0.6,0.114)
                    (0.8,0.106)
                    (1.0,0.122)
                };
                \addlegendentry{RDA}
        
                \addplot[
                    color=red,
                    mark=*,
                    thick
                ] coordinates {
                    (0.0,0.100)
                    (0.2,0.101)
                    (0.4,0.103)
                    (0.6,0.106)
                    (0.8,0.113)
                    (1.0,0.132)
                };
                \addlegendentry{SA-CN}
        
                \addplot[
                    color=IGABlue,   
                    mark=triangle*,
                    thick
                ] coordinates {
                    (0.0,0.100)
                    (0.2,0.152)
                    (0.4,0.189)
                    (0.6,0.210)
                    (0.8,0.223)
                    (1.0,0.238)
                };
                \addlegendentry{IGA-LWP}
        
            \end{axis}
            \end{tikzpicture}
        \caption{C.\ elegans}  
        \label{fig:sub-celegans}
    \end{subfigure}

    \vspace{0.5em}

    \begin{subfigure}[t]{0.48\textwidth}
        \centering
           \begin{tikzpicture}
                \begin{axis}[
                    width=\textwidth,
                    height=0.75\textwidth,
                    xlabel={Proportion},
                    ylabel={RMSE},
                    xmin=0, xmax=1.0,
                    ymin=0.06, ymax=0.16,
                    enlarge x limits=0.03,   
                    xtick={0,0.2,0.4,0.6,0.8,1.0},
                    ytick={0.06,0.08,0.10,0.12,0.14,0.16},
                    scaled y ticks=false,
                    yticklabel style={
                        /pgf/number format/fixed,
                        /pgf/number format/precision=2
                    },
                     legend style={
                        draw=black,
                        fill=white,
                        at={(0.03,0.97)},
                        anchor=north west
                    },
                    axis lines=box,
                    tick align=inside,      
                    font=\scriptsize,
                    legend cell align=left
                ]
            
                    \addplot[
                        color=black,
                        mark=square*,
                        thick
                    ] coordinates {
                        (0.0,0.07)
                        (0.2,0.068)
                        (0.4,0.074)
                        (0.6,0.086)
                        (0.8,0.087)
                        (1.0,0.087)
                    };
                    \addlegendentry{RDA}
            
                    \addplot[
                        color=red,
                        mark=*,
                        thick
                    ] coordinates {
                        (0.0,0.07)
                        (0.2,0.071)
                        (0.4,0.068)
                        (0.6,0.096)
                        (0.8,0.099)
                        (1.0,0.102)
                    };
                    \addlegendentry{SA-CN}
            
                    \addplot[
                        color=IGABlue,   
                        mark=triangle*,
                        thick
                    ] coordinates {
                        (0.0,0.07)
                        (0.2,0.071)
                        (0.4,0.091)
                        (0.6,0.117)
                        (0.8,0.126)
                        (1.0,0.156)
                    };
                    \addlegendentry{IGA-LWP}
            
                \end{axis}
                \end{tikzpicture}
        \caption{Netscience}
        \label{fig:sub-netscience}
    \end{subfigure}
    \hfill
    \begin{subfigure}[t]{0.48\textwidth}
        \centering
            \begin{tikzpicture}
                    \begin{axis}[
                        width=\textwidth,
                        height=0.75\textwidth,
                        xlabel={Proportion},
                        ylabel={RMSE},
                        xmin=0, xmax=1.0,
                        ymin=0.18, ymax=0.38,
                        enlarge x limits=0.03,   
                        xtick={0,0.2,0.4,0.6,0.8,1.0},
                        ytick={0.18,0.20,0.22,0.24,0.26,0.28,0.30,0.32,0.34,0.36,0.38},
                        scaled y ticks=false,
                        yticklabel style={
                            /pgf/number format/fixed,
                            /pgf/number format/precision=2
                        },
                         legend style={
                            draw=black,
                            fill=white,
                            at={(0.5,0.68)},
                            anchor=north west
                        },
                        axis lines=box,
                        tick align=inside,      
                        font=\scriptsize,
                        legend cell align=left
                    ]
                
                        \addplot[
                            color=black,
                            mark=square*,
                            thick
                        ] coordinates {
                            (0.0,0.188)
                            (0.2,0.189)
                            (0.4,0.201)
                            (0.6,0.203)
                            (0.8,0.203)
                            (1.0,0.204)
                        };
                        \addlegendentry{RDA}
                
                        \addplot[
                            color=red,
                            mark=*,
                            thick
                        ] coordinates {
                            (0.0,0.188)
                            (0.2,0.190)
                            (0.4,0.200)
                            (0.6,0.203)
                            (0.8,0.206)
                            (1.0,0.214)
                        };
                        \addlegendentry{SA-CN}
                
                        \addplot[
                            color=IGABlue,   
                            mark=triangle*,
                            thick
                        ] coordinates {
                            (0.0,0.188)
                            (0.2,0.241)
                            (0.4,0.301)
                            (0.6,0.338)
                            (0.8,0.352)
                            (1.0,0.372)
                        };
                        \addlegendentry{IGA-LWP}
                
                    \end{axis}
                    \end{tikzpicture}
        
        \caption{UC-net}
        \label{fig:sub-ucnet}
    \end{subfigure}

    \caption{RMSE vs. perturbation proportion for different attack methods on various datasets.}
    \label{fig:four-datasets}
\end{figure}

Under the local attack setting, we investigate how the RMSE metric changes with the perturbation scale, as shown in Fig. \ref{fig:four-datasets}. The perturbation scale is quantified as the ratio of the number of perturbed links to the degree of the target link. The experimental results indicate that, as the perturbation ratio increases, the attack effectiveness of IGA-LWP consistently improves, whereas the improvements for RDA and SA-CN are much slower and even negligible on some datasets. Due to its inherent randomness,  RDA fails to effectively capture how perturbations should be adjusted. Although   SA-CN  may be useful as a reference for link prediction in social networks, its effectiveness is limited in other types of networks, such as biological networks, where the common-neighbor index is invalid.  In contrast, IGA-LWP generates perturbations based on gradient information and adjusts link weights along the gradient direction, enabling it to accurately capture the optimal direction for weight perturbation; consequently, its attack performance continues to improve significantly as the perturbation ratio increases.

\subsection{Transferability to DeepWalk, Node2Vec and GCN}

Finally, we evaluate the transferability of IGA-LWP. We generate adversarial graphs under a local attack setting with a perturbation ratio of $0.5k_t$, using IGA-LWP, RDA, and SA-CN respectively. Then, we perform link weight prediction with DeepWalk, Node2Vec, and GCN on both the original and adversarially perturbed graphs, measuring performance using RMSE. The results in Fig. 3 demonstrate that adversarial graphs crafted by IGA-LWP consistently cause the highest RMSE across all three prediction models and datasets, while RDA and SA-CN yield much weaker impacts. This suggests that SEA effectively captures critical structural features of the graph, and perturbations guided by SEA gradients remain potent against different link weight prediction methods.Therefore, IGA-LWP produces adversarial graphs that are highly effective and demonstrate notable cross-model transferability.
\begin{figure}[htbp]
    \centering
    \begin{subfigure}[t]{0.95\textwidth} 
        \centering
            \begin{tikzpicture}
                \begin{axis}[
                ybar,
                ybar legend,
                width=0.85\textwidth,
                height=0.40\textwidth,
                title={Deepwalk},
                title style={font=\bfseries},
                xlabel={Datasets},
                ylabel={RMSE},
                xlabel style={font=\small},
                ylabel style={font=\small},
                ymin=0, ymax=0.30,
                ytick={0,0.1,0.2,0.3},
                yticklabel style={font=\scriptsize},
                xticklabel style={font=\scriptsize},
                symbolic x coords={Neural-net, C.\ elegans, Netscience, UC-net},
                xtick=data,
                bar width=8pt,
                enlarge x limits=0.18,        
                axis lines=box,
                tick align=inside,
                ymajorgrids=true,
                major grid style={dashed,gray!35},
                legend style={
                    draw=black!70,
                    fill=white,
                    at={(0.5,0.85)},
                    anchor=south,
                    legend columns=4,
                    font=\scriptsize
                },
                legend cell align=left,
                legend image code/.code={
                    \draw[#1] (0cm,-0.10cm) rectangle (0.30cm,0.10cm);
                }   
            ]
        
                \addplot+[
                    bar shift=-18pt,
                    fill=OrigColor!90,
                    draw=black!40
                ] coordinates {
                    (Neural-net,0.22)
                    (C.\ elegans,0.145)
                    (Netscience,0.12)
                    (UC-net,0.215)
                };
                \addlegendentry{original}
        
                \addplot+[
                    bar shift=-10pt,
                    fill=RDAColor!95,
                    draw=black!40
                ] coordinates {
                    (Neural-net,0.23)
                    (C.\ elegans,0.150)
                    (Netscience,0.13)
                    (UC-net,0.225)
                };
                \addlegendentry{RDA}
        
                \addplot+[
                    bar shift=-2pt,
                    fill=SACNColor!95,
                    draw=black!40
                ] coordinates {
                    (Neural-net,0.24)
                    (C.\ elegans,0.152)
                    (Netscience,0.14)
                    (UC-net,0.235)
                };
                \addlegendentry{SA-CN}
        
                \addplot+[
                    bar shift=6pt,
                    fill=IGABlue!75,
                    draw=black!40
                ] coordinates {
                    (Neural-net,0.25)
                    (C.\ elegans,0.18)
                    (Netscience,0.145)
                    (UC-net,0.245)
                };
                \addlegendentry{IGA-LWP}
        
            \end{axis}
            \end{tikzpicture}
            \label{fig:sub-a}
        \end{subfigure}
    \vspace{0.1em} 

    \begin{subfigure}[t]{0.95\textwidth}
        \centering
        \begin{tikzpicture}
        \begin{axis}[
            ybar,
            ybar legend,
            width=0.85\textwidth,
            height=0.40\textwidth,
            title={Node2vec},
            title style={font=\bfseries},
            xlabel={Datasets},
            ylabel={RMSE},
            xlabel style={font=\small},
            ylabel style={font=\small},
            ymin=0, ymax=0.30,
            ytick={0,0.1,0.2,0.3},
            yticklabel style={font=\scriptsize},
            xticklabel style={font=\scriptsize},
            symbolic x coords={Neural-net, C.\ elegans, Netscience, UC-net},
            xtick=data,
            bar width=8pt,
            enlarge x limits=0.18,         
            axis lines=box,
            tick align=inside,
            ymajorgrids=true,
            major grid style={dashed,gray!35},
            legend style={
                draw=black!70,
                fill=white,
                at={(0.5,0.85)},
                anchor=south,
                legend columns=4,
                font=\scriptsize
            },
            legend cell align=left,
            legend image code/.code={
                \draw[#1] (0cm,-0.10cm) rectangle (0.30cm,0.10cm);
            }   
        ]
    
            \addplot+[
                bar shift=-18pt,
                fill=OrigColor!90,
                draw=black!40
            ] coordinates {
                (Neural-net,0.222)
                (C.\ elegans,0.155)
                (Netscience,0.125)
                (UC-net,0.245)
            };
            \addlegendentry{original}
    
            \addplot+[
                bar shift=-10pt,
                fill=RDAColor!95,
                draw=black!40
            ] coordinates {
                (Neural-net,0.230)
                (C.\ elegans,0.160)
                (Netscience,0.13)
                (UC-net,0.255)
            };
            \addlegendentry{RDA}
    
            \addplot+[
                bar shift=-2pt,
                fill=SACNColor!95,
                draw=black!40
            ] coordinates {
                (Neural-net,0.240)
                (C.\ elegans,0.165)
                (Netscience,0.132)
                (UC-net,0.265)
            };
            \addlegendentry{SA-CN}
    
            \addplot+[
                bar shift=6pt,
                fill=IGABlue!75,
                draw=black!40
            ] coordinates {
                (Neural-net,0.248)
                (C.\ elegans,0.175)
                (Netscience,0.145)
                (UC-net,0.278)
            };
            \addlegendentry{IGA-LWP}
    
        \end{axis}
        \end{tikzpicture}
        \label{fig:sub-b}
    \end{subfigure}
    \vspace{0.1em}

    \begin{subfigure}[t]{0.95\textwidth}
        \centering
        \begin{tikzpicture}
        \begin{axis}[
            ybar,
            ybar legend,
            width=0.85\textwidth,
            height=0.40\textwidth,
            title={GCN},
            title style={font=\bfseries},
            xlabel={Datasets},
            ylabel={RMSE},
            xlabel style={font=\small},
            ylabel style={font=\small},
            ymin=0, ymax=0.30,
            ytick={0,0.1,0.2,0.3},
            yticklabel style={font=\scriptsize},
            xticklabel style={font=\scriptsize},
            symbolic x coords={Neural-net, C.\ elegans, Netscience, UC-net},
            xtick=data,
            bar width=8pt,
            enlarge x limits=0.18,        
            axis lines=box,
            tick align=inside,
            ymajorgrids=true,
            major grid style={dashed,gray!35},
            legend style={
                draw=black!70,
                fill=white,
                at={(0.5,0.85)},
                anchor=south,
                legend columns=4,
                font=\scriptsize
            },
            legend cell align=left,
            legend image code/.code={
                \draw[#1] (0cm,-0.10cm) rectangle (0.30cm,0.10cm);
            }   
        ]
    
            \addplot+[
                bar shift=-18pt,
                fill=OrigColor!90,
                draw=black!40
            ] coordinates {
                (Neural-net,0.222)
                (C.\ elegans,0.155)
                (Netscience,0.125)
                (UC-net,0.245)
            };
            \addlegendentry{original}
    
            \addplot+[
                bar shift=-10pt,
                fill=RDAColor!95,
                draw=black!40
            ] coordinates {
                (Neural-net,0.230)
                (C.\ elegans,0.160)
                (Netscience,0.13)
                (UC-net,0.255)
            };
            \addlegendentry{RDA}
    
            \addplot+[
                bar shift=-2pt,
                fill=SACNColor!95,
                draw=black!40
            ] coordinates {
                (Neural-net,0.240)
                (C.\ elegans,0.165)
                (Netscience,0.132)
                (UC-net,0.265)
            };
            \addlegendentry{SA-CN}
    
            \addplot+[
                bar shift=6pt,
                fill=IGABlue!75,
                draw=black!40
            ] coordinates {
                (Neural-net,0.248)
                (C.\ elegans,0.175)
                (Netscience,0.145)
                (UC-net,0.278)
            };
            \addlegendentry{IGA-LWP}
        \end{axis}
        \end{tikzpicture}
        \label{fig:sub-b}
    \end{subfigure}
    \caption{RMSE of the link weight prediction methods (Deepwalk, Node2vec, and GCN) on adversarial graphs generated by different attack methods.}
\end{figure}
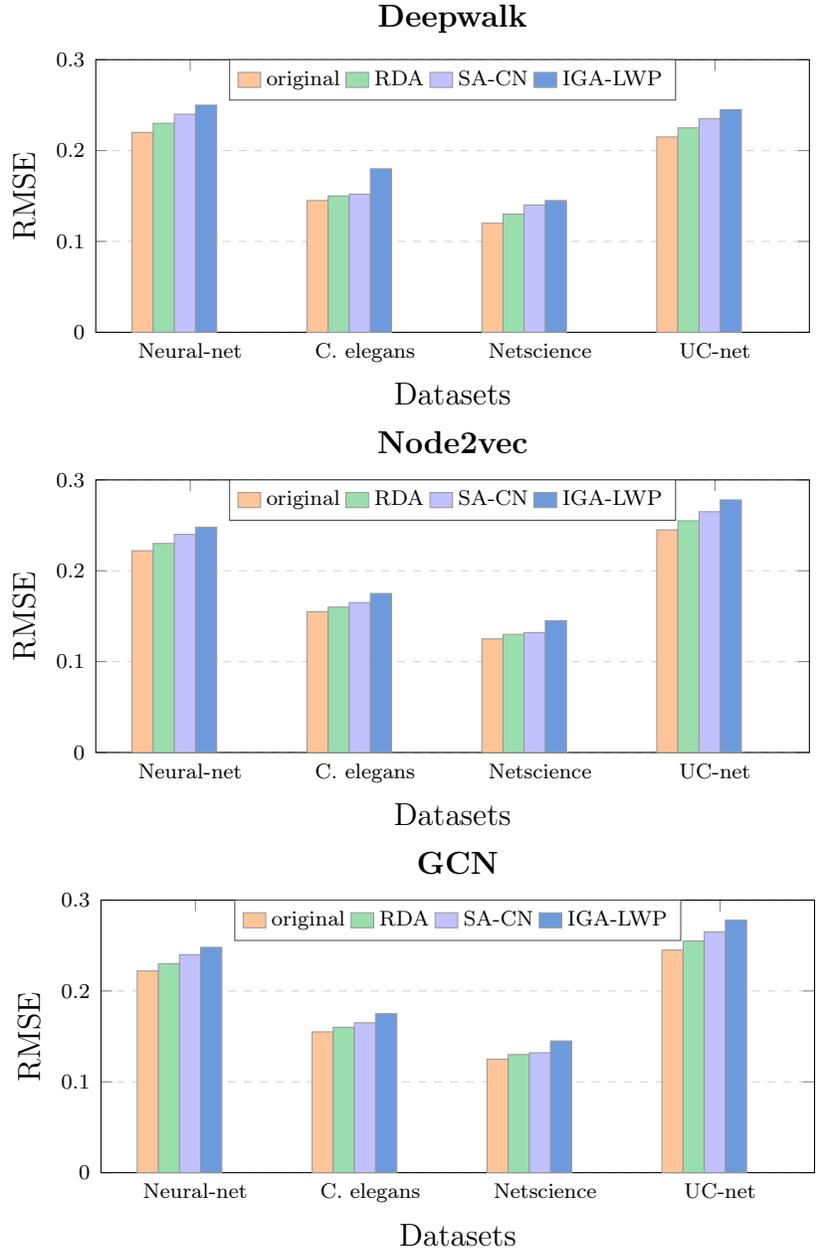

\section{Conclusion}
\label{sec_conclusion}
In summary, we study adversarial attacks on link weight prediction in complex networks. We propose IGA-LWP, an iterative gradient-based attack method designed based on the prediction model SEA. Experiments on four real-world weighted networks demonstrate that IGA-LWP can effectively attack various link weight prediction methods: by adding only small-scale perturbations to the link weights, it can significantly decrease the performance of multiple link weight prediction models. Therefore, IGA-LWP can be used both as a tool for privacy protection and as an evaluation method for assessing the robustness of link weight prediction models.


\section*{Acknowledgements}
This work was supported in part by the National Natural Science Foundation of China under Grant 62427808.

\bibliographystyle{elsarticle-num-names} 
\bibliography{refs}

\end{document}